\documentclass[final,11pt,3p,sort&compress]{elsarticle} 
\makeatletter
\def\ps@pprintTitle{%
 \let\@oddhead\@empty
 \let\@evenhead\@empty
 \def\@oddfoot{\reset@font\hfil\thepage\hfil}%
 \let\@evenfoot\@oddfoot}

\biboptions{super}	

\usepackage[T1]{fontenc} 
\usepackage[latin2]{inputenc}
\usepackage{amssymb}
\usepackage{amsthm}

\usepackage{graphicx,cancel}
\usepackage{ulem}
\usepackage{color}
\usepackage{fixltx2e} 
\usepackage{microtype} 
\usepackage[
   separate-uncertainty  = true,
   mode = math,
   multi-part-units      = brackets,
]{siunitx} 
\usepackage{hyperref}
\hypersetup{
    colorlinks=true, 
    linkcolor=blue,  
    citecolor=blue,  
}
\usepackage[all]{hypcap}  

\renewcommand{\_}[1]{{}_{\mathrm{#1}}}

\begin{document}

\begin{frontmatter}

\title{\textbf{Interplay of hydrogen bonding and molecule--substrate interaction in self-assembled adlayer structures of a hydroxyphenyl-substituted porphyrin}}

\author[afko]{Lars~Smykalla}
\author[afko]{Pavel~Shukrynau}
\address[afko]{Technische Universität Chemnitz, Institute of Physics, Solid Surfaces Analysis Group, D-09107 Chemnitz, Germany}
\author[chem]{Carola~Mende}
\author[chem]{Tobias~Rüffer}
\author[chem]{Heinrich~Lang}
\address[chem]{Technische Universität Chemnitz, Institute of Chemistry, Inorganic Chemistry, D-09107 Chemnitz, Germany}
\author[afko]{Michael~Hietschold}

\date{\today}

\begin{abstract}

The formation of hydrogen-bonded organic nano-structures and the role of the substrate lattice thereby were investigated by scanning tunneling microscopy. The self-organization of 5,10,15,20-tetra(\textit{p}-hydroxyphenyl)porphyrin (H$\_2$THPP) molecules leads to two molecular arrangements on Au(111). One of these is characterized by pair-wise hydrogen bonding between hydroxyl groups and a low packing density which enables a rotation of individual molecules in the structure. A different interaction with stronger chain-like hydrogen bonding and additional interactions of phenyl groups was observed for the second structure. The influence of the substrate on the epitaxial behavior is demonstrated by the adsorption of H$\_2$THPP on the highly anisotropic Ag(110) substrate. There, several balances between the occupation of favorable adsorption positions and the number of hydrogen bonds per molecule were found.
The molecules form molecular chains on Ag(110) and also assemble into two-dimensional periodic arrangements of differently sized close-packed blocks similar to the second type of supramolecular ordering found on Au(111). Dispersion corrected Density Functional Theory calculations were applied to understand the adsorption and complex epitaxy of these molecules. It is shown that the azimuthal orientation of the saddle-shape deformed molecule plays an important role not only for the intermolecular but also for the molecule--substrate interaction.

\end{abstract}

\end{frontmatter}

\section{Introduction}

The bottom-up fabrication of molecular nano-devices with advanced functions is a subject of significant interest.\cite{Stepanow2008, Hietschold2005, Yoshimoto2010} The construction of desired molecular network architectures on suitable substrates is achieved by the self-assembly of molecular building blocks, which is mainly controlled by a complex interplay between non-covalent bonds such as hydrogen bonds, van der Waals and electrostatic interactions, and the epitaxy on the substrate.\cite{Smykalla2012, Scheffler2013, Hipps2002, Scudiero2003, Hill2006a, Meier2005, Gopakumar2008, Toader2010, Toader2011, Tian2010, Rojas2012} From the knowledge gained about the delicate balance between intermolecular and molecule--substrate interactions, the functional nano-networks can then be tuned by adjusting functional groups of the molecule to gain the desired architectures\cite{Yokoyama2004, Fendt2009, Heim2010} and the interface electronic structure for the specific application. Therefore, it is important that the molecular orientation, configuration and conformation, and the interactions between the adsorbed molecules and the substrate as well as between molecules can be identified and understood, e.g. by using scanning tunneling microscopy (STM).

Porphyrins and their appropriate metal complexes are a very important class of molecules, which are responsible for the functionality in many biological systems.
The stability and exceptional versatility of porphyrins led to great interest in the self-assembly of these molecules on surfaces for nano-technology.\cite{Gottfried2009}
Numerous studies demonstrate that porphyrin-based molecules tend to arrange themselves in well-defined long-range ordered lateral structures.\cite{Yoshimoto2003, Auwaerter2006, Auwaerter2010, Rojas2010, Beggan2012} The combination of both functionality and structure makes them ideal candidates for molecular devices with applications, such as for electronics, spintronics\cite{Wende2007}, data storage\cite{THPPspec}, non-linear optics~\cite{Ogawa2002}, sensors\cite{Rakow2000}, solar cells\cite{Mozer2009}, catalysts\cite{Fukuzumi2001}, as well as for tumor photosensitizers and photodynamic therapy\cite{Berenbaum1986, Bonnett1989, Ibrahim2011}.
In this study, we use hydroxyphenyl functional groups at the meso positions of a porphyrin molecule to enhance the intermolecular interaction and self-assembly properties \textit{via} hydrogen bonding\cite{Lu2009, Bhyrappa1997, Lei2001, Hill2007, Garcia-Lekue2012}.

For hydrogen-bonded networks, often an incommensurate registry between the lattices of the close-packed molecular layer and the substrate can be expected. Weak intermolecular interactions and large energetic differences in adsorption sites likely result in a commensurate epitaxy or coincidence with a commensurate super cell.\cite{Hooks2001} On the other hand, the prediction of the specific mode represents a problem if the molecular arrangement is determined by strong intermolecular interactions but also the substrate plays a major role in the epitaxy.
The mechanisms by which the molecular self-assemblies compensate for the incommensurateness of the lattice parameters of their close-packed crystal structures are then of particular interest.
Au(111) was chosen as a surface with a flat and Ag(110) as one with an highly anisotropic and corrugated surface potential.
Results for the novel molecular structures formed on these substrates and the investigation of the interactions involved in the self-organization of the molecules are presented in the following.

\section{Methods}
\subsection{Experimental details}

5,10,15,20-tetra(\textit{p}-hydroxy\-phenyl)\-porphyrin (H$\_2$THPP) was synthesized analogous to a published procedure.\cite{Adler1967, THPPspec} Clean surfaces of the Au(111) and Ag(110) single crystals were prepared by multiple cycles of Ar$^+$ sputtering at an energy of \SI{500}{eV} and annealing to \SI{400}{\celsius} for \SI{1}{\hour}. H$\_2$THPP molecules were deposited on the substrate by organic molecular beam epitaxy in ultra high vacuum (UHV). For this, solid H$\_2$THPP was first purified by heating to a temperature slightly below the sublimation temperature in UHV. The molecules were then deposited at around \SI{350}{\celsius} on the clean Au(111) and Ag(110) surfaces. The temperature of the substrate during deposition was kept at room temperature. The scanning tunneling microscopy experiments were performed with a variable-temperature STM from Omicron in UHV. The base pressure in the UHV chamber was in the range of \SI{E-10}{mbar}. Electrochemically etched tungsten tips were used for the STM. All measurements were done at room temperature (ca. \SI{23}{\celsius}). STM images were measured in constant current mode with a tunneling current of \SI{100}{pA} unless specified otherwise. The bias voltage as referred to in the figure caption was applied to the sample. STM data was processed with the WSxM software~\cite{Horcas2007} whereby moderate low-pass filtering was used for reduction of noise.

\subsection{Computational details}

Density Functional Theory (DFT) calculations were performed with the grid-based projector augmented wave method (GPAW)\cite{Enkovaara2010}. The revised version of the exchange-correlation functional of Perdew, Burke and Ernzerhof (RPBE)~\cite{Hammer1999} was used in the calculations for different adsorption sites and energy barriers for rotation of H$\_2$THPP on Ag(110). For the dispersive interactions not accounted for in DFT~\cite{Klimes2012}, the correction from Tkatchenko and Scheffler\cite{Tkatchenko2009} (vdW(TS)) was added, which was reported to give good adsorption distances and energies.\cite{Marom2011} The proposed vdW-parameters by Ruiz \textit{et al.} were used, which include screening effects of the metal surfaces.\cite{Ruiz2012}
The substrate was modeled by a slab of 4 layers and ($6\times 8$) size in $x$ and $y$ for Ag(110) and ($7\times 8$) size for Au(111) with periodic boundary conditions in the $x$ and $y$ directions and zero boundary condition in $z$ direction. An \SI{8}{\angstrom} wide vacuum region was added in $z$ direction above the adsorbed molecule and \SI{6}{\angstrom} vacuum under the substrate slab. The computed lattice constant of \SI{4.17}{\angstrom} was used for Au and \SI{4.09}{\angstrom} for Ag. Due to its large periodicity, the reconstruction of the Au(111) surface was not considered within our calculations. For the Brillouin zone sampling, only the $\Gamma$-point was used due to the relative big size of the cell. The topmost substrate layer and all atoms of the molecule were allowed to relax during optimization.

We applied the LCAO mode\cite{Larsen2009} of GPAW, where the pseudo density is evaluated on a fine grid, whereas a double-$\zeta$ plus valence polarization type basis set of atomic orbitals is used for the wave functions. For localized basis sets, the overlap of the atomic orbitals used for the substrate with those used as basis set for the molecular adsorbate results in a decrease of the total energy, which can lead to a severe artificial overestimation of the adsorption energy, especially for molecules on metallic surfaces.\cite{Buimaga-Iarinca2014} This basis-set superposition error (BSSE) increases with decreasing molecule--substrate separation and, thus, was estimated by the counterpoise (CP) procedure for each geometry.\cite{Boys1970} It should be noted that counterpoise-corrected energies are often higher than the binding energy at the limit of a complete basis set, where the BSSE vanishes, due to the additional error from the incompleteness of the basis set.\cite{Halkier1999} For the basis set used, the value of the correction is, for example, around 35\% of the uncorrected vdW-DF2 binding energy for H$\_2$THPP on Au(111) at the distance of the energetic minimum.
The adsorption energy $E\_{ads}$ was computed by: $E\_{ads} = E\_{mol+subs} - E\_{mol}^{CP} - E\_{subs}^{CP} - E\_{mol}^{def} - E\_{subs}^{def}$.
Thereby, $E\_{mol}^{CP}$ and $E\_{subs}^{CP}$ are the counterpoise-corrected energies of the individual molecule and substrate, respectively, at the geometry in the combined molecule-surface system; $E\_{mol}^{def}$ and $E\_{subs}^{def}$ are the corresponding deformation energies relative to the relaxed, isolated geometries.
In the calculations of the binding energy as a function of adsorption height, the optimized molecular geometry of the fully relaxed H$\_2$THPP molecule on Au(111) was used and fixed for all distances. The binding energy does not contain the deformation energy.

\section{Results and Discussion}

\subsection{Adlayer structures of H$\_2$THPP on Au(111)}

\begin{figure}[htb]
\centering
\includegraphics[width=0.45\textwidth]{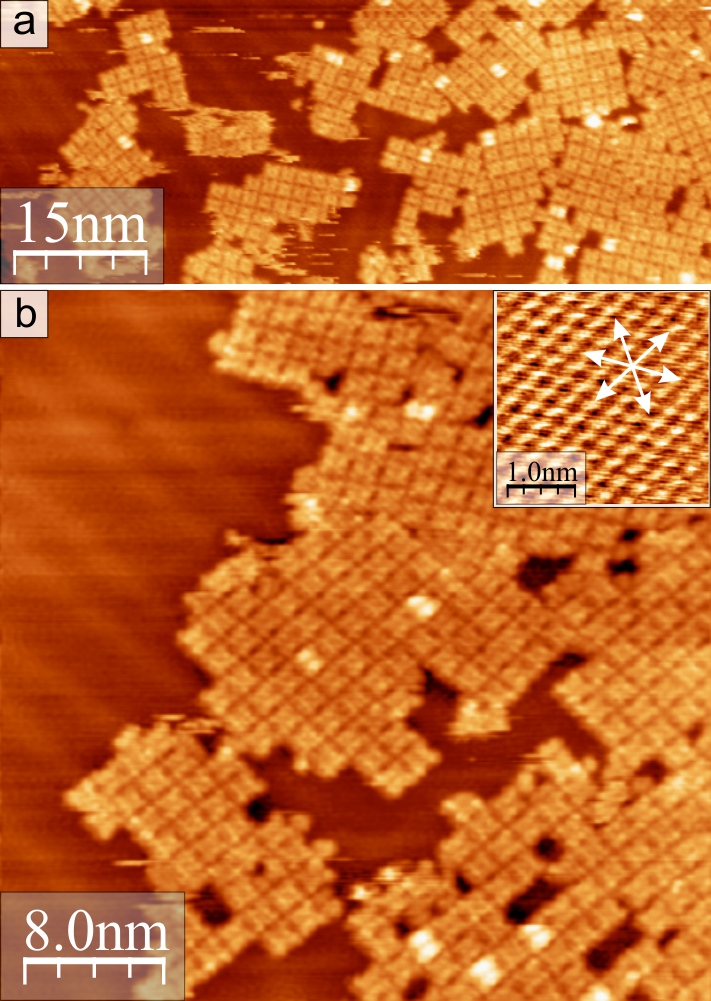}
\caption{(a),(b) Formation of small islands at submonolayer coverage. Molecules on top of the first layer appear brighter ($U = -\SI{1}{V}$, $I=\SI{100}{pA}$). Inset in (b): atomically resolved Au lattice ($U = -\SI{0.5}{V}$, $I=\SI{500}{pA}$). Arrows are along the direction of the unit cell vectors of the Au(111) substrate.}
\label{fig:subML}
\end{figure}

\begin{figure}[!tb]
\centering
\includegraphics[width=0.75\textwidth]{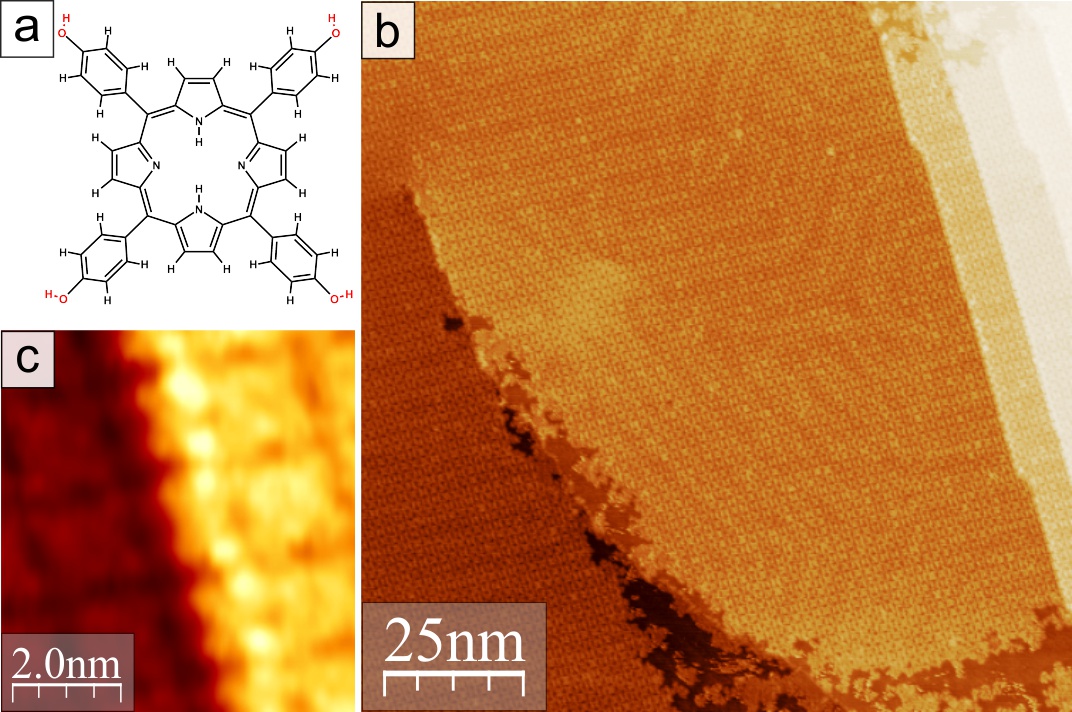}
\caption{(a) Chemical structure of H$\_2$THPP. (b) Large ordered domain of H$\_2$THPP on Au(111) after annealing to $\approx \SI{150}{\celsius}$ ($U = -\SI{1.5}{V}$). (c) Magnification on H$\_2$THPP molecules adsorbed on a mono-atomic step edge of the Au substrate ($U = -\SI{1.5}{V}$).}
\label{fig:ML}
\end{figure}

\begin{figure*}[!bt]
\centering
\includegraphics[width=0.99\textwidth]{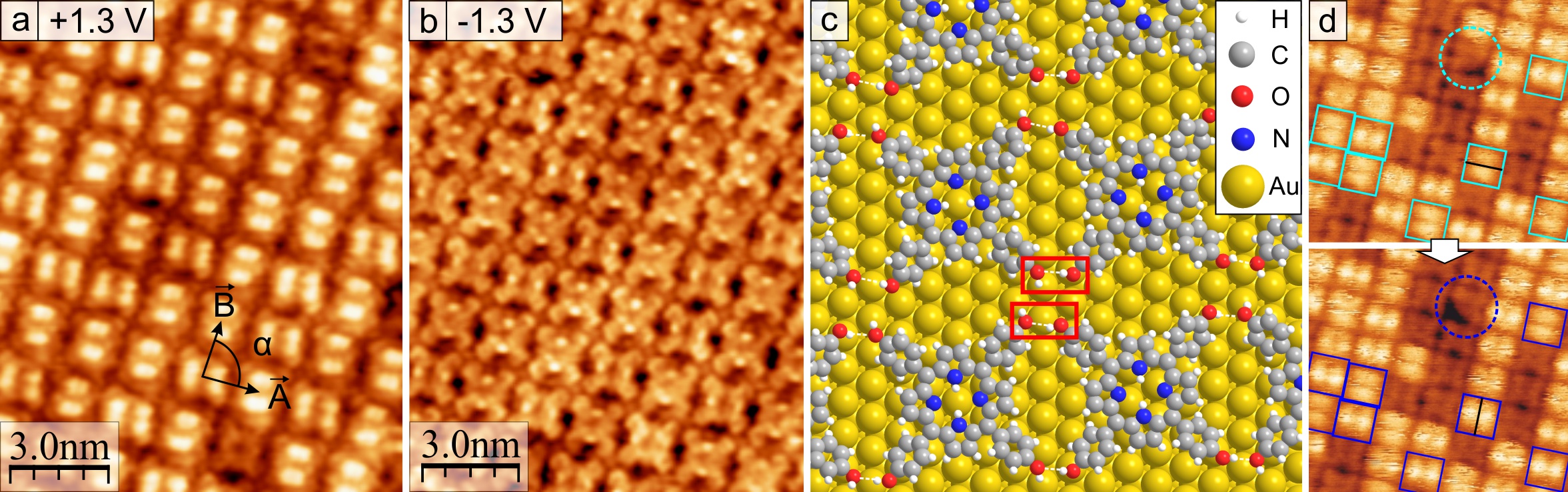}
\caption{Highly resolved filled (a) and empty molecular states (b) STM image of the square structure of H$\_2$THPP on Au(111). (c) Model for the arrangement of H$\_2$THPP molecules in this structure. Pair-wise hydrogen bonding is marked with red rectangles. The reconstruction of Au(111) is not included in the model of the epitaxy. (d) Molecules which are rotated by \SI{90}{\degree} in the subsequently measured image (bottom) are marked by squares. A rotated molecule with one missing phenyl group is indicated by a dotted circle ($\SI{11}{nm} \times \SI{10.5}{nm}$, $U = +\SI{1.2}{V}$, sample annealed to $\approx \SI{150}{\celsius}$).}
\label{fig:struc1}
\end{figure*}

Upon adsorption of H$\_2$THPP molecules on the Au(111) surface, the molecules begin to arrange in small islands. As can be seen in Fig.~\ref{fig:subML}, at a coverage of around 0.8 of a molecular monolayer, these islands have a structure with a square molecular unit cell and domains where, with respect to each other, the lattice is rotated by \SI{120}{\degree} due to the threefold symmetry of the Au(111) surface. The domain walls between fcc and hcp regions of Au (pairs of corrugation lines) from the $22\times\sqrt{3}$ reconstruction of Au(111) are visible in the left part of Fig.~\hyperref[fig:subML]{\ref*{fig:subML}(b)}. The reconstruction is visible through the first molecular monolayer [Fig.~\hyperref[fig:ML]{\ref*{fig:ML}(b)}], which is an indication for physisorption of the molecules. At room temperature, the molecules are mobile on the surface [small streaks from movement while scanning in Fig.~\hyperref[fig:subML]{\ref*{fig:subML}(a)}] and the rearrangement and growth of the small domains can be followed with STM. Furthermore, a few individual H$\_2$THPP molecules adsorbed on top of an underlying molecule [Fig.~\hyperref[fig:subML]{\ref*{fig:subML}(a)}].
After the sample was annealed at around \SI{150}{\celsius}, highly ordered areas with a very large domain size are found [Fig.~\hyperref[fig:ML]{\ref*{fig:ML}(b)}]. Monolayer islands of this structure expand continuously over atomic step edges of the Au(111) substrate with the centers of the tilted H$\_2$THPP molecules located directly over the step edge [Fig.~\hyperref[fig:ML]{\ref*{fig:ML}(c)}]. One unit cell vector of the molecular structure is parallel to one lattice vector of Au(111). A highly resolved STM image of the square molecular arrangement is displayed in Fig.~\ref{fig:struc1}. The lattice parameters measured with STM for the square structure are in average $|\vec{A}| = \SI{1.74 \pm 0.04}{nm}$, $|\vec{B}| = \SI{1.64\pm 0.04}{nm}$ and the angle between both unit cell vectors is $\alpha = \SI{87\pm 1}{\degree}$. In the error margin of the measured values a point-on-line epitaxy~\cite{Hooks2001} on Au(111) can be found, where H$\_2$THPP adsorbs commensurately on identical adsorption positions along $\vec{A} = 6\cdot \vec{a} \approxeq \SI{1.732}{nm}$ [Fig.~\hyperref[fig:struc1]{\ref*{fig:struc1}(c)}].
The herringbone $22\times\sqrt{3}$ reconstruction of Au(111) is neglected in the proposed epitaxy, however, the flexible molecular structure should be able to adapt to the local small displacements of the underlying Au atoms. In this rectangular arrangement of H$\_2$THPP, hydroxyl groups of adjacent molecules form likely O$-$H$\cdots$O hydrogen bonding pairs [marked in Fig.~\hyperref[fig:struc1]{\ref*{fig:struc1}(c)}]. The reason why this structure does not show notable preferential growth along the direction of the hydrogen bonding ($\vec{B}$) could be due to the aforementioned likely commensurate and, therefore, slightly more favorable adsorption along the other lattice direction ($\vec{A}$).

The appearance of the molecules is different for positive [empty states, Fig.~\hyperref[fig:struc1]{\ref*{fig:struc1}(a)}] and negative bias voltage [filled states, Fig.~\hyperref[fig:struc1]{\ref*{fig:struc1}(a)}], respectively. The general shape of the H$\_2$THPP molecule with the four tilted phenyl groups and the depression in the center is seen at negative bias voltage. At positive voltage, H$\_2$THPP is imaged as two pairs of protrusions which originate from $\pi$ electrons of the pyrrolic carbon atoms on a saddle-shaped porphyrin macrocycle. It is known that 5,10,15,20-tetraphenylporphyrin (TPP) molecules adsorbed on a metal surface show a non-planar deformation of the central porphyrin unit.\cite{Hill2007, Auwaerter2010, Diller2012} It is characterized by the bending of pairs of opposite pyrrole rings above and below the porphyrin mean plane, respectively, to form a saddle-shape. This conformation is induced by the steric repulsion from the phenyl rings, which are rotated toward a more parallel adsorption geometry to increase the interaction of the $\pi$ electrons with the surface.
A mirror plane defined by the molecular saddle-shape is rotated by $\delta_B = \SI{4 \pm 2}{\degree}$ relative to $\vec{B}$. Within the molecular layer and also on top of step edges of Au(111) individual molecules can be rotated randomly by \SI{90}{\degree}.
This is possible because in this structure the intermolecular interaction is dominated by hydrogen bonding and does not include a more rigid $\pi$-$\pi$ interaction\cite{Sinnokrot2004,Grimme2008}. In the case of metal- and free-base TPP molecules on (111) metal surfaces, the edge-to-face configuration of adjacent phenyl groups leads to a fixed orientation of all molecules.\cite{Yoshimoto2003, Auwaerter2010, Beggan2012} An azimuthal \SI{90}{\degree} rotation of individual H$\_2$THPP molecules in the molecular network can be occasionally induced by the tip while scanning, which is demonstrated in Fig.~\hyperref[fig:struc1]{\ref*{fig:struc1}(d)}. If the two successively measured images are compared with each other, correspondingly rotated molecules can be identified (marked with squares). This change is indeed a rotation of the full molecule in the structure and not a flipping of the saddle-shape deformation as evidenced by the rotation of a defective molecule with one missing phenyl group, which is marked with a circle in Fig.~\hyperref[fig:struc1]{\ref*{fig:struc1}(d)}. This indicates a higher degree of freedom for the individual molecules in the structure which is stabilized by hydrogen bonds between hydroxyl groups compared to an arrangement with $\pi$-$\pi$ interaction. The two different appearances of this molecule at higher bias voltages, which become evident from Fig.~\hyperref[fig:struc1]{\ref*{fig:struc1}(d)}, and the correlated electronic structure are discussed elsewhere.\cite{THPPspec} 

\begin{figure}[!tb]
\centering
\includegraphics[width=0.75\textwidth]{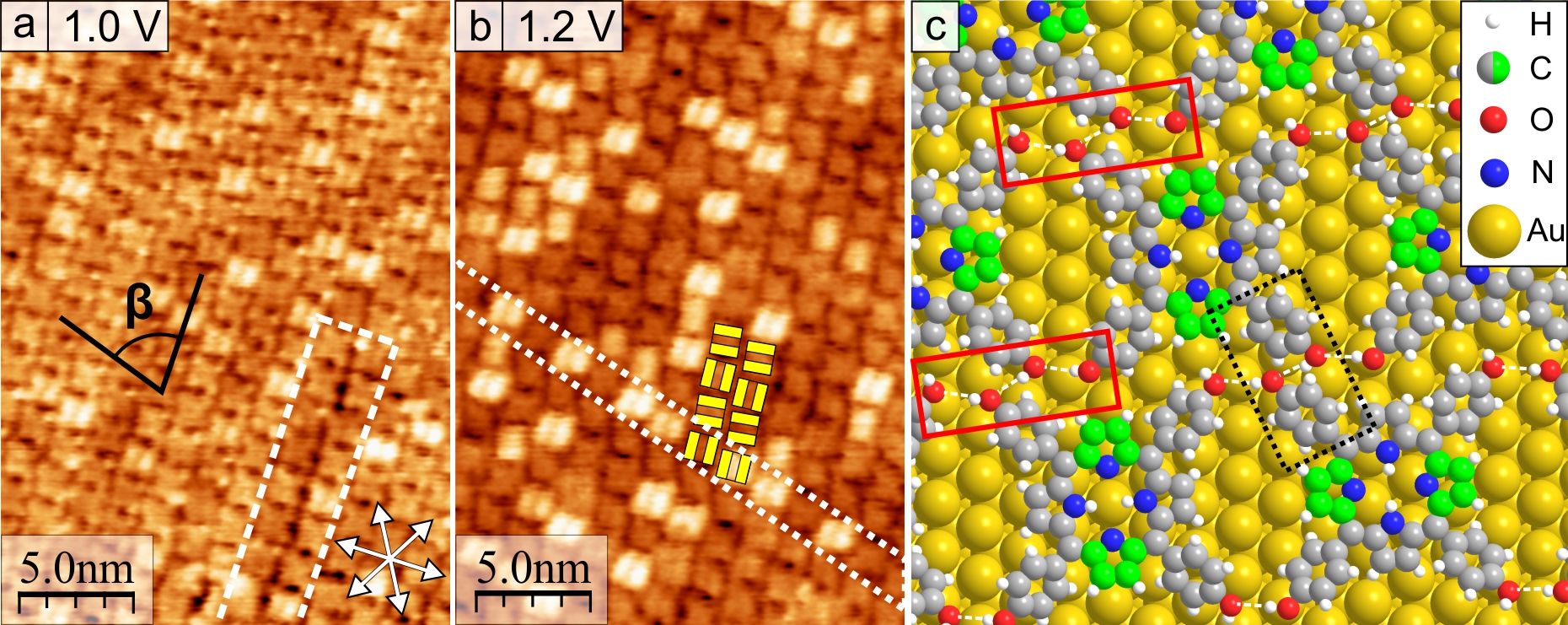}
\caption{STM images of the oblique structure of H$\_2$THPP on Au(111): (a) The lines of the $22\times\sqrt{3}$ reconstruction of Au are weakly visible through the molecular monolayer. A lattice defect by the square arrangement is marked. (b) At higher bias voltage the saddle-shape of H$\_2$THPP (indicated by yellow rectangles) can be identified -- along the direction of one lattice vector molecules have identical and along the other vector alternating orientations (sample annealed to $\approx \SI{200}{\celsius}$). (c) Model of the molecular arrangement. The interaction of the phenyl groups is marked exemplarily in black, hydrogen bonding in red and the protruding C atoms of the saddle-shaped molecule in green. The reconstruction of Au(111) is not included in the model of the epitaxy.}
\label{fig:struc2}
\end{figure}

At submonolayer coverages, also after annealing, only the square structure was found. A second type of arrangement of H$\_2$THPP molecules was observed on the Au(111) surface together with the aforementioned structure for coverages close to one complete monolayer. The unit cell is oblique with an angle of $\beta = \SI{70 \pm 2}{\degree}$ (Fig.~\ref{fig:struc2}). The packing density of this array is $(0.42 \pm 0.04)$ molecules per nm$^2$ and clearly larger than that of the square structure with $(0.35 \pm 0.02)$ molecules per nm$^2$. The epitaxial angle of the adsorbate lattice with respect to the Au(111) lattice is around \SI{20}{\degree}, which was measured by using the double lines of the reconstruction of Au. The marked area in Fig.~\hyperref[fig:struc2]{\ref*{fig:struc2}(a)} is a lattice defect in the form of a small domain of the square structure, which demonstrates that one lattice vector of both structures is identical. Fig.~\hyperref[fig:struc2]{\ref*{fig:struc2}(b)} shows the oblique structure at higher bias voltage where the saddle-shape of individual molecules can be seen. All molecules along one lattice vector (dotted lines) have an identical orientation of the saddle-shaped macrocycle [$\SI{24 \pm 4}{\degree}$ relative to this vector], whereas along the other lattice vector the molecules are rotated alternately by circa \SI{90}{\degree}. Molecules in every second row appear less resolved in Fig.~\hyperref[fig:struc2]{\ref*{fig:struc2}(b)} due to the orientation of the saddle-shape nearly vertical to the fast scanning direction (clearly resolved two protrusions if the scanning direction is rotated by \SI{90}{\degree}).
The model for the structure and its coincident epitaxy on Au(111) is presented in Fig.~\hyperref[fig:struc2]{\ref*{fig:struc2}(c)}. For the molecular arrangement in the unit cell with an angle of \SI{70}{\degree}, hydrogen bonding is always zigzag-chain-wise between four hydroxyl groups. Compared to the square structure, this increases the number of hydrogen bonds per molecule from two to three. Thus, the overall intermolecular interaction in the oblique structure is higher. A similar triple hydrogen bonding scheme was reported before for 5,10,15,20-tetrakis(3,5-dimethyl-4-hydroxyphenyl)porphyrin on Cu(111).\cite{Hill2007} Moreover, for the crystal structure of this molecule Hill \textit{et al.} found arrays of hydrogen-bonded molecules, whose bonding scheme resembles that of the square structure in our system.\cite{Hill2007} In the oblique arrangement of H$\_2$THPP, the planes of the close phenyl groups are always parallel to each other (black, dotted rectangle in Fig.~\hyperref[fig:struc2]{\ref*{fig:struc2}(c)}). We propose that this is due an additional favorable interaction between the parallel-displaced hydroxyphenyl groups.\cite{Sinnokrot2004, Seo2009} Thus, this leads to the row-wise and rigid azimuthal \SI{90}{\degree} alternation of the saddle-shaped H$\_2$THPP molecules. A clearly defined fixed orientation was also observed for TPP with an identical rotation of all molecules within the molecular structure, which was explained by the edge-to-face stacking of the tilted phenyl rings.\cite{Auwaerter2010, Beggan2012}

\subsection{Adlayer structures of H$\_2$THPP on Ag(110)}

\begin{figure}[!tb]
\centering
\includegraphics[width=0.5\textwidth]{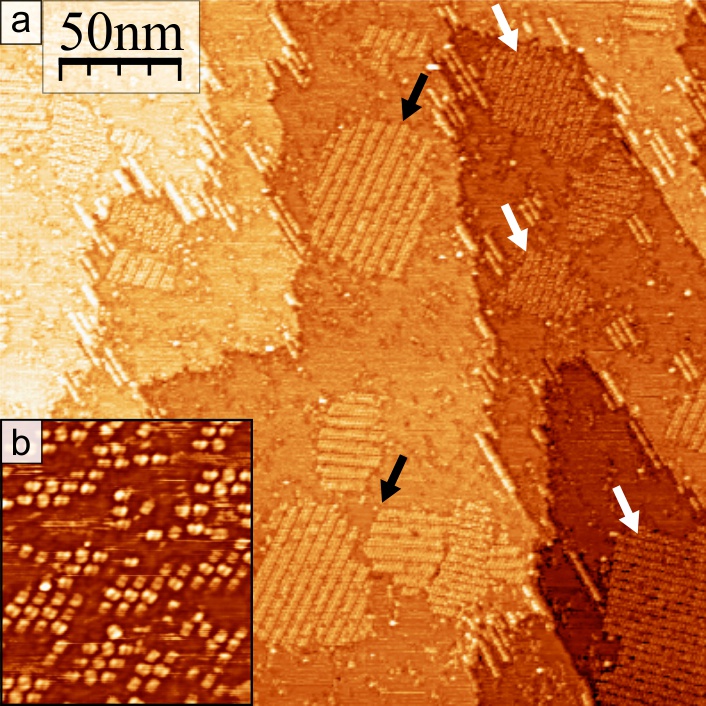}
\caption{(a) Large scale STM image of submonolayer coverage showing molecular chains and ordered islands of H$\_2$THPP on Ag(110) after annealing to $\approx \SI{200}{\celsius}$ (black/white arrows for the different 2D structures, $U = +\SI{1.0}{V}$). (b) Aggregation into small molecular islands after the first annealing to $\approx \SI{150}{\celsius}$ ($\SI{23.1}{nm} \times \SI{26.2}{nm}$, $U = +\SI{1.4}{V}$).}
\label{fig:overview_Ag110}
\end{figure}

To acquire knowledge about the influence of the corrugation of the surface potential on the epitaxy and supramolecular ordering, the adsorption of H$\_2$THPP on the highly anisotropic Ag(110) substrate was also investigated.
Immediately after deposition of a submonolayer coverage, most molecules on the surface were disordered (not shown) and single molecules could be stably imaged at room temperature, which indicates a lower diffusion activity of H$\_2$THPP molecules on Ag(110) than on Au(111). Post-annealing of the sample to \SI{150}{\celsius} increased the mobility, resulting in the formation of many small ordered islands on the substrate terraces [Fig.~\hyperref[fig:overview_Ag110]{\ref*{fig:overview_Ag110}(b)}]. A typical molecular island consisted of nine or twelve molecules. During heating to circa \SI{200}{\celsius}, the molecules rapidly diffused and rearranged into molecular chains, striped islands and differently ordered areas, which are shown in Fig.~\hyperref[fig:overview_Ag110]{\ref*{fig:overview_Ag110}(a)}. After annealing of the sample to even higher temperatures, no indication of an influence from possibly deprotonated $-$OH groups on the molecular structures was found by STM.

\begin{figure*}[!htb]
\centering
\includegraphics[width=0.99\textwidth]{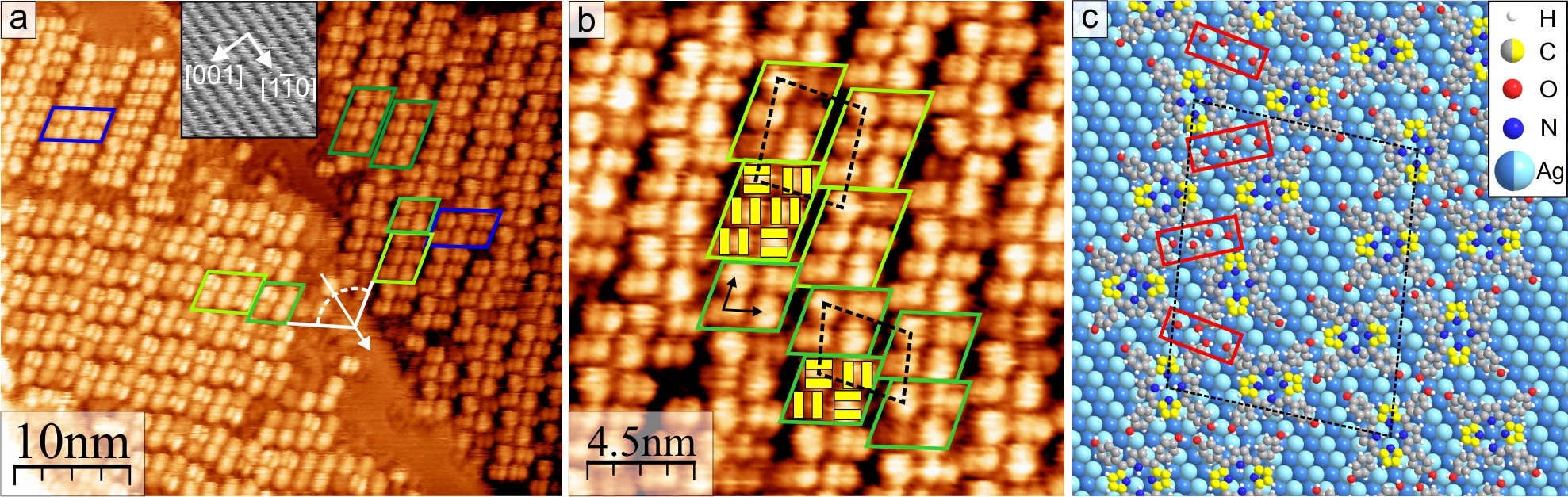}
\caption{(a) Stripe-like structure (blue) and mirrored domains of the mixed structures consisting of close-packed blocks (green parallelograms) containing four, six or eight molecules ($U = +\SI{1.0}{V}$, $I=\SI{100}{pA}$, annealed to $\approx \SI{200}{\celsius}$). Inset: resolved atomic rows of the Ag(110) substrate ($\SI{4}{nm} \times \SI{4}{nm}$, $U = +\SI{0.5}{V}$, $I=\SI{1}{nA}$). (b) Magnification of the periodic-block structure of H$\_2$THPP ($U = +\SI{1.0}{V}$). Unit cells are indicated by the black dashed parallelograms.
(c) Model for the arrangement of the molecules in the structure and the epitaxy on Ag(110). To emphasize the orientation, protruding carbon atoms in the macrocycle are colored yellow [yellow rectangles in (b)]. First Ag layer bright blue, second layer dark blue. Sites of hydrogen bonding are marked exemplarily with red rectangles.}
\label{fig:struktur_Ag110}
\end{figure*}

Each structure can occur mirrored relative to the $[1\bar{1}0]$ direction of Ag(110) resulting in two possible domains, as shown in Fig.~\hyperref[fig:struktur_Ag110]{\ref*{fig:struktur_Ag110}(a)}.
All observed two dimensional structures of H$\_2$THPP on Ag(110) are in principle related because they are based on a supramolecular arrangement very similar to the structure with an angle of the unit cell of $\SI{70\pm 2}{\degree}$ of H$\_2$THPP on Au(111). But unlike the adsorption on Au(111), the molecular assemblies on Ag(110) were not continuously close-packed.
One structure is characterized by periodically arranged close-packed ``blocks'' containing either four ($\approx$ 25\% of island area), six (67\%) or seldom eight or more (8\%) molecules  (framed in green in Fig.~\ref{fig:struktur_Ag110}).
These blocks often occurred mixed because one vector of the unit cells is identical and only the length of the blocks differs. The model of this structure is shown in Fig.~\hyperref[fig:struktur_Ag110]{\ref*{fig:struktur_Ag110}(c)}, where the unit cell and also the chain-wise O$-$H$\cdots$O bonding between the molecules (red rectangles) is marked.
The intermolecular interactions in a block increase with its size but also each shows hydrogen bonds with adjacent blocks. In total, this leads to 1.5 hydrogen bonds per molecule independent of the extent of the close-packed arrangement due to the proportionally sized unit cells.
The molecules in each block have well defined, rigid orientations of the saddle-shape as marked in yellow in Fig.~\hyperref[fig:struktur_Ag110]{\ref*{fig:struktur_Ag110}(b),(c)}.

\begin{figure}[!tb]
\centering
\includegraphics[width=0.65\textwidth]{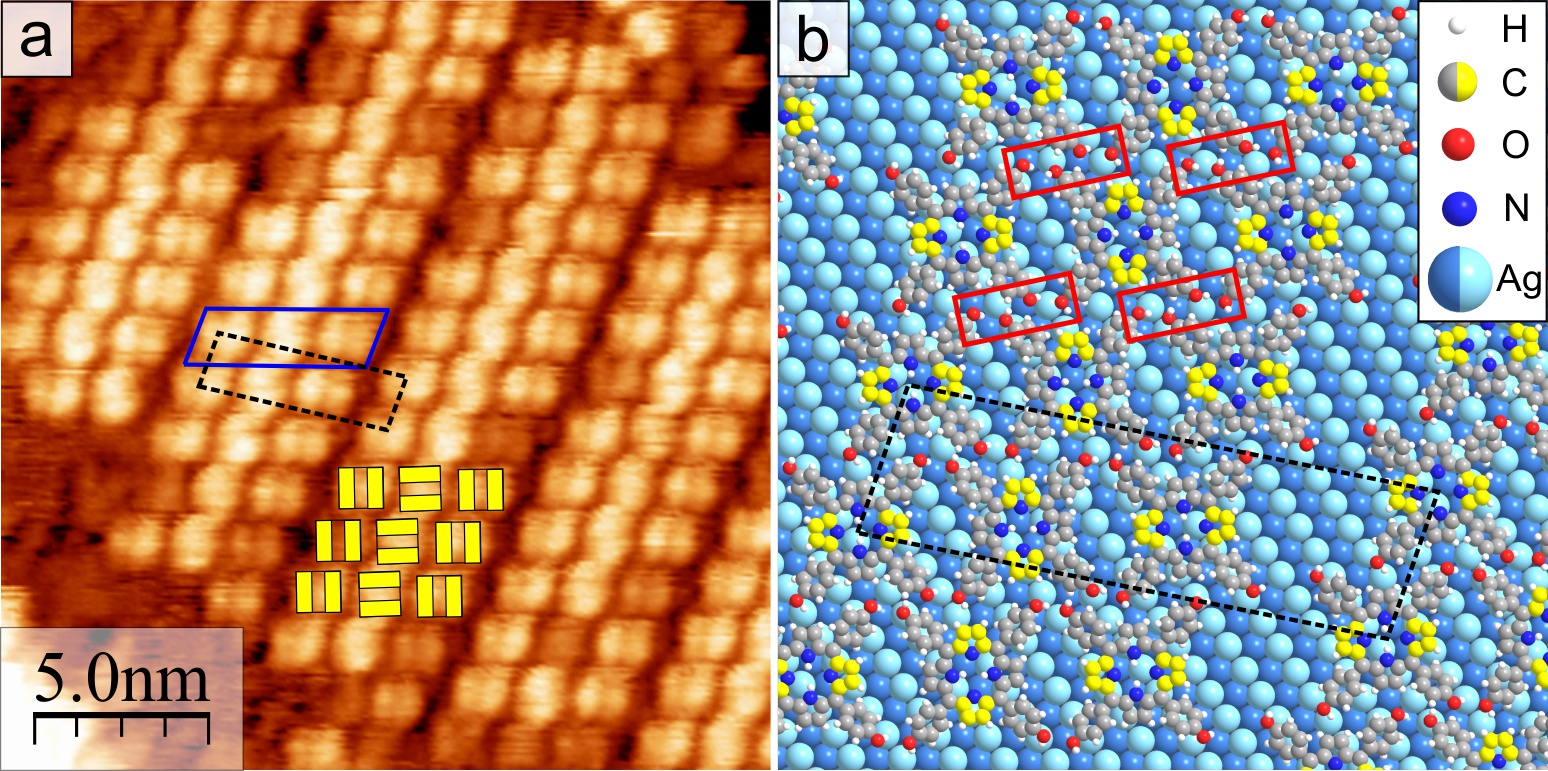}
\caption{(a) Stripe-like structure of H$\_2$THPP on Ag(110) ($U = +\SI{1.0}{V}$, annealed to $\approx \SI{200}{\celsius}$). The unit cell (dashed, black) contains an arrangement of three molecules (framed in blue). (b) Model of epitaxy of this structure. Carbon atoms in the pyrrole rings bended away from the surface colored in yellow, Ag atom in first layer in bright blue, in second layer dark blue.}
\label{fig:streifen_Ag110}
\end{figure}

The second type of structure (Fig.~\ref{fig:streifen_Ag110}) can be characterized as stripes, each with a width of three molecules. The molecular saddle-shape is rotated by \SI{90}{\degree} for molecules located diagonally (top left to bottom right in Fig.~\ref{fig:streifen_Ag110}) likely again due to an interaction of the hydroxyphenyl groups, which leads to the alternating azimuthal rotation for rows of H$\_2$THPP inside a stripe. The molecular stripe is rotated by $\approx \SI{53}{\degree}$ relative to $[1\bar{1}0]$ of Ag(110). Correspondingly, the molecules are rotated by $\varepsilon_1 = \SI{30\pm 7}{\degree}$ and $\varepsilon_2 = \SI{-60\pm 7}{\degree}$ relative to $[1\bar{1}0]$ [see Fig.~\hyperref[fig:DFT_Ag110]{\ref*{fig:DFT_Ag110}(a)}], which is also valid for H$\_2$THPP in the blocks of the other structure.
The model of the stripe-like structure is shown in Fig.~\hyperref[fig:streifen_Ag110]{\ref*{fig:streifen_Ag110}(b)}, where the unit cell, which contains three molecules, is marked by a dashed rectangle. The molecular arrangement is, as previously mentioned, identical to the oblique structure on Au(111). The striking difference to the periodic-block array is the continued close-packing along one direction and a width of three instead of two molecules as in the block arrangements. This leads to an increase of the intermolecular interaction with in average two hydrogen bonds per molecule. The packing density of the stripe-like structure is with $(0.38 \pm 0.03)$ molecules per nm$^2$ not different from those of the block pattern with unit cells of four and six molecules. But, they are all clearly smaller than the packing density of the previously discussed close-packed oblique structure on Au(111) due to the spacing between the molecular stripes or blocks. On our samples, the stripe-like structure was the most often observed one, which indicates that it is energetically slightly more favorable than the periodic-block array.

\begin{figure}[!tb]
\centering
\includegraphics[width=0.65\textwidth]{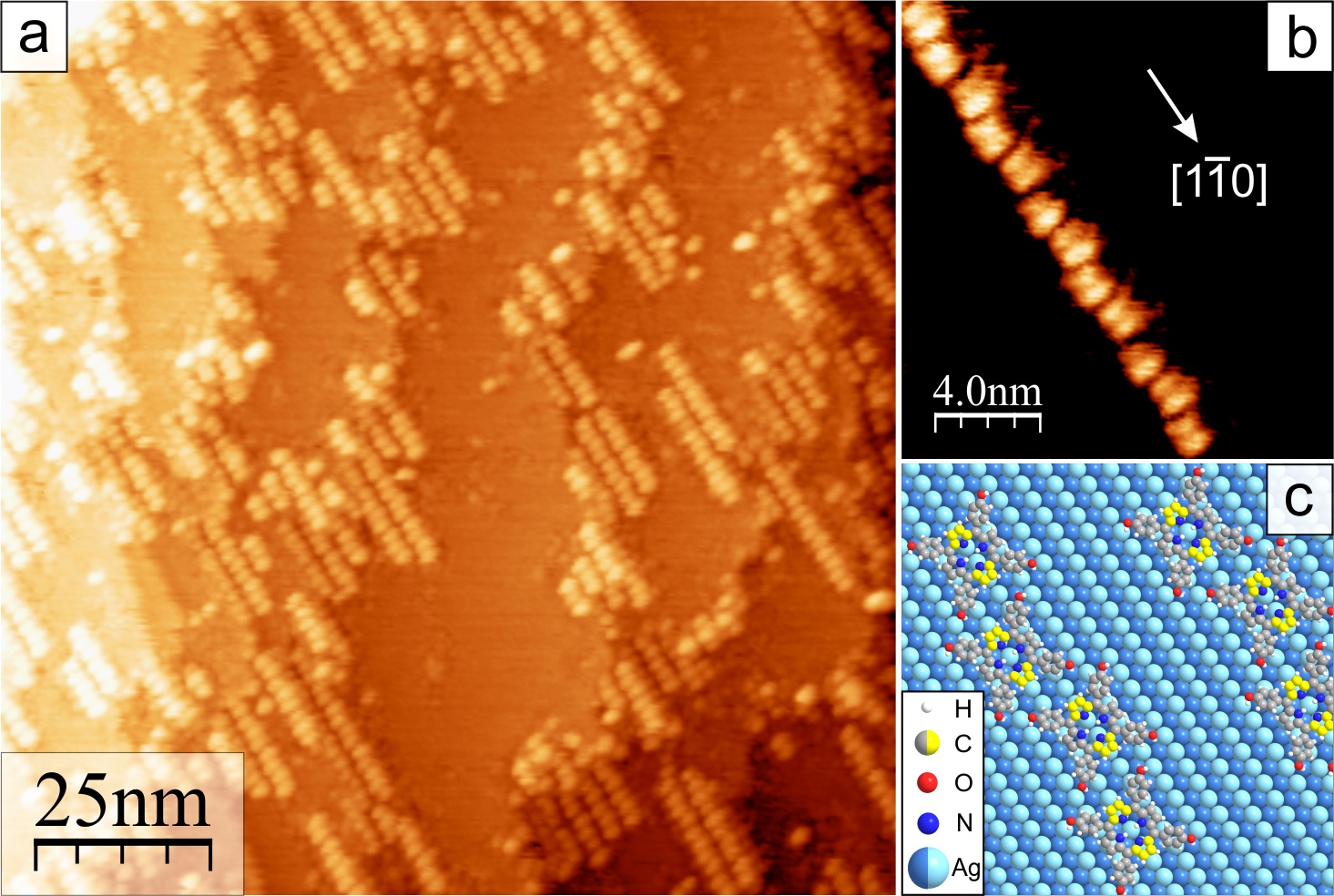}
\caption{(a) Molecular chains along $[1\bar{1}0]$ near step edges of Ag(110) ($U = +\SI{0.7}{V}$). (b) Magnification of a chain. H$\_2$THPP molecules are slightly displaced in a zigzag manner ($U = +\SI{1.6}{V}$). (b) Model of epitaxy of the molecules in chains on Ag(110). Carbon atoms in the pyrrole rings bended away from the surface colored in yellow, Ag atom in first layer in bright blue, in second layer dark blue.}
\label{fig:chain}
\end{figure}

One-dimensional chains of H$\_2$THPP were formed along the $[1\bar{1}0]$ direction of Ag(110) as shown in Fig.~\ref{fig:chain}. These chains had, at the investigated coverage, an average length of around \SI{12}{nm}. In most cases, they were located near steps of the Ag(110) surface likely because step edges are often favorable adsorption sites and starting points for further growth. The lower diffusion barrier along $[1\bar{1}0]$ compared to $[001]$ could then promote the assembly into chains. A closer look at this molecular arrangement reveals that the molecules are slightly displaced in a zigzag manner and oriented with the saddle shape parallel to $[1\bar{1}0]$ [Fig.~\hyperref[fig:chain]{\ref*{fig:chain}(b)}]. The distance between the molecules is $\SI{1.66 \pm 0.05}{nm}$ and vertically between neighboring chains $\SI{3.2 \pm 0.2}{nm}$. The corresponding structural model is shown in Fig.~\hyperref[fig:chain]{\ref*{fig:chain}(c)}.

The determination of the exact adsorption sites in the epitaxy of the molecular structures was inaccessible to STM imaging.
However, knowledge about this is necessary to understand the influence of the Ag(110) lattice and to clarify the reason for the different supramolecular assemblies. In the following section, DFT calculations for the adsorption of H$\_2$THPP on Au(111) and Ag(110) at different positions and rotational angles of the molecule relative to the substrate lattice will be presented and the implications for the energetic balances, responsible for the structures, be discussed.

\subsection{Calculation of the adsorption energies and discussion}

\begin{figure}[!tb]
\centering
\includegraphics[width=0.65\textwidth]{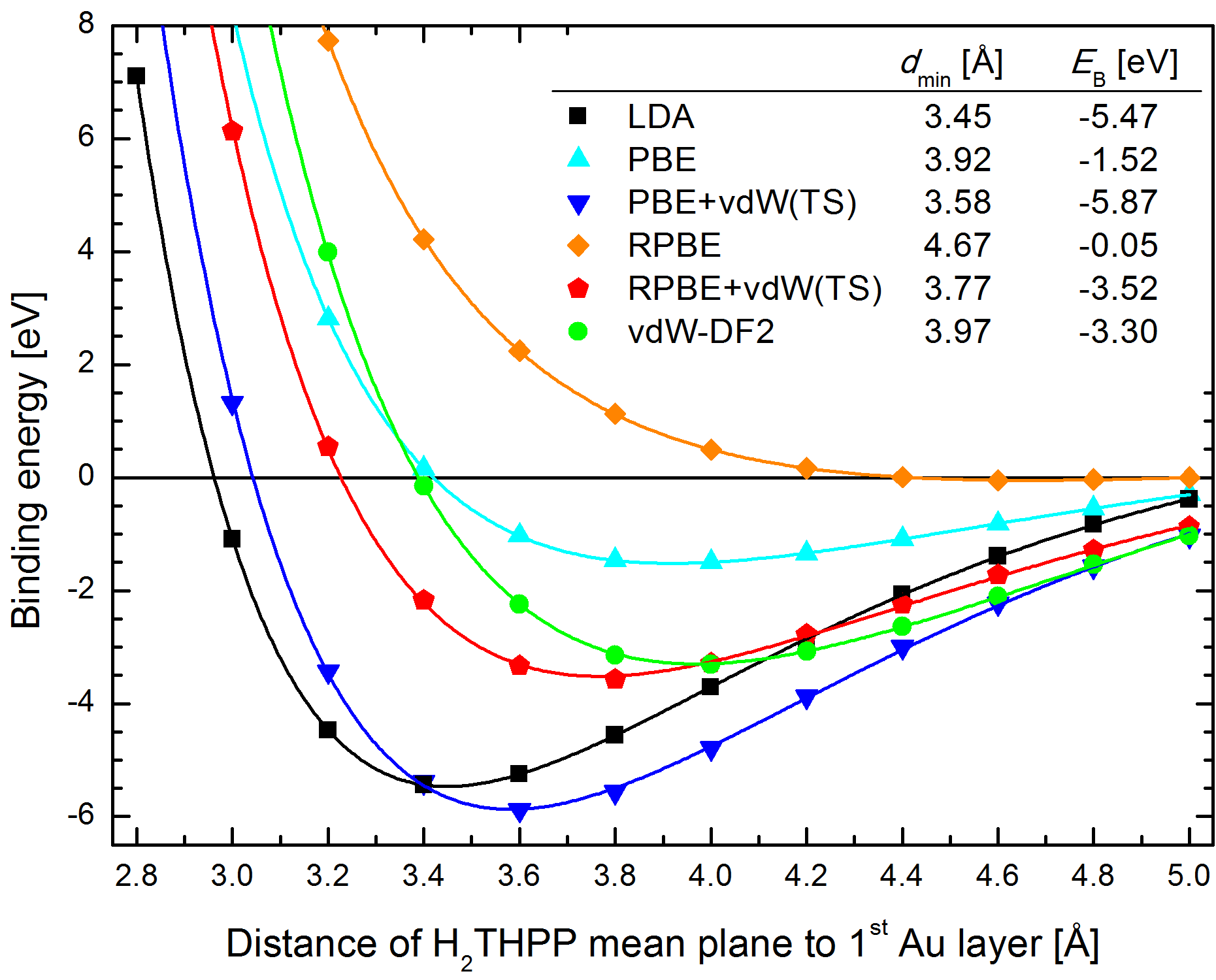}
\caption{Binding energies as a function of vertical distance $d$ between the molecular plane of H$\_2$THPP and the Au(111) surface for different exchange-correlation functionals and dispersion corrections. The values of the energetic minima are summarized in the inserted table.}
\label{fig:THPP_Au_dist_33}
\end{figure}

First of all, it should be noted that for DFT the adsorption height and energy in general depend strongly on the approximations used. Therefore, for the investigated system several exchange-correlation functionals were tested (Fig.~\ref{fig:THPP_Au_dist_33}). It is often problematic to estimate which functional performs the best because the experimental adsorption energy and molecule--substrate separation are only available for few, mostly small molecules.

The RPBE functional is a revised version of PBE to improve atomization and chemisorption energies for small molecules.\cite{Hammer1999} It can be clearly seen that RPBE leads to almost no binding for physisorption like in the case of H$\_2$THPP on Au(111). The missing non-local interaction can be compensated by adding the dispersion correction by Tkatchenko and Scheffler~\cite{Tkatchenko2009, Ruiz2012} [vdW(TS)] with the van der Waals parameters for metal surfaces by Ruiz \textit{et al}.\cite{Ruiz2012} PBE+vdW(TS) gives a smaller equilibrium adsorption height $d\_{min}$ compared to PBE but the binding energy for H$\_2$THPP on Au(111) becomes then similar to that for the local density approximation (LDA). In general, LDA is well known to overestimate binding while many xc-functionals of the generalized gradient approximation result in an underestimation. At the moment, one of the more advanced functionals which include vdW interaction self-consistently is vdW-DF2.\cite{Lee2010} This non-local correlation functional is without the need of predefined dispersion coefficients and was shown to give very good binding energies, although adsorption distances are still too large.\cite{Klimes2012,Buimaga-Iarinca2014} Fig.~\ref{fig:THPP_Au_dist_33} shows that when using RPBE with the vdW(TS) correction instead of PBE, the binding energy is now close to the value calculated with vdW-DF2 and additionally the adsorption height is improved. It should also be noted that the computational cost for RPBE+vdW(TS) is significantly lower compared to vdW-DF2. From these comparisons, we conclude that RPBE+vdW(TS) should give the best description for both energy and geometry of our systems.

The optimized geometry of H$\_2$THPP on the Au(111) surface is a saddle-shaped deformation with angles of the phenyl planes of $\phi\_{ph, ad} = \SI{32.5}{\degree}$ and pyrrole planes of $\phi\_{pyr, ad} = \SI{23}{\degree}$ relative to the surface plane. This geometry leads to the almost same distance of the carbon atoms closest to the Au surface in the phenyl and the pyrrole rings. $E\_{ads}-d$ curves were also calculated for the fixed geometry of H$\_2$THPP optimized in gas phase with $\phi\_{ph, gas} = \SI{65}{\degree}$ and $\phi\_{pyr, gas}= \SI{7}{\degree}$. This leads to an adsorption height higher by circa \SI{0.15}{\angstrom} and an absolute value of the binding energy smaller by \SI{1.1}{eV} [for RPBE+vdW(TS)]. Thus, this gain in binding energy justifies the $\approx \SI{0.45}{eV}$ necessary for the deformation from a planar to a saddle-shaped macrocycle. Different adsorption sites on Au(111) and angles of azimuthal rotation relative to the substrate lattice were considered. Only a small difference of ca. \SI{0.05}{eV} was found between all of them, which indicates small energy barriers for diffusion and rotation of H$\_2$THPP on Au(111) and, therefore, moderate mobility at room temperature (kinetic energy of \SI{0.025}{eV}). A similar calculated diffusion barrier of \SI{0.032}{eV} has been reported for H$\_2$TPP on Ag(111).\cite{Rojas2010}

\begin{figure}[!tb]
\centering
\includegraphics[width=0.90\textwidth]{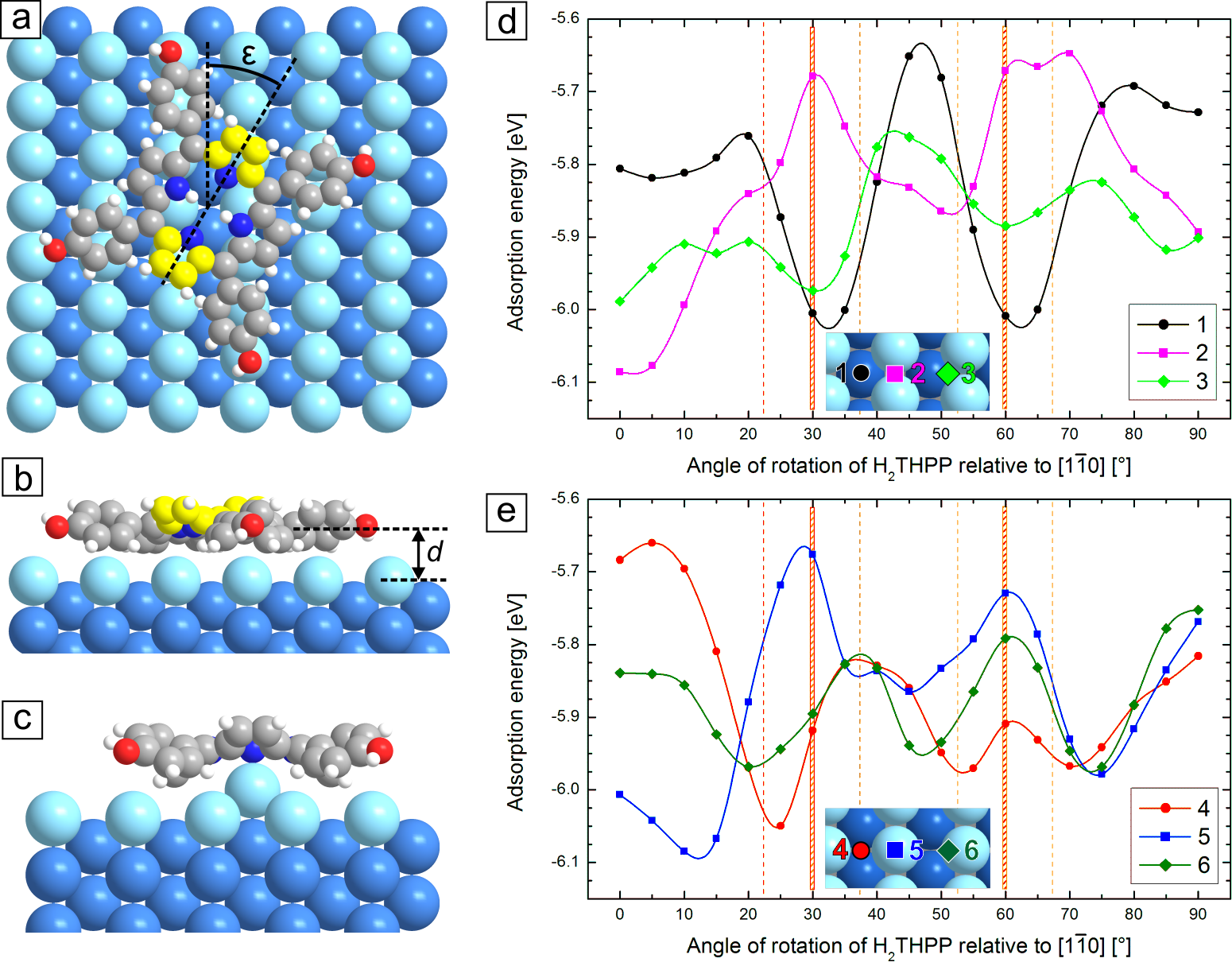}
\caption{(a) Model of the adsorption geometry of H$\_2$THPP on Ag(110). The angle of azimuthal rotation $\varepsilon$ of the molecule relative to $[1\bar{1}0]$ is indicated. Investigated adsorption sites on the Ag(110) lattice are marked and numbered. The two pyrrole rings tilted away from the surface are colored yellow. (b) Side view showing the adsorption geometry at $\varepsilon = \SI{30}{\degree}$ and definition of the adsorption height $d$. (c) Model for the adsorption on an additional protruding row of Ag adatoms. (d),(e) Adsorption energy in dependence of the azimuthal rotation of H$\_2$THPP for the indicated adsorption positions of the molecular center at a constant height of $d=\SI{3.2}{\angstrom}$. The connecting splines are only a guide for the eye.}
\label{fig:DFT_Ag110}
\end{figure}

Fig.~\hyperref[fig:DFT_Ag110]{\ref*{fig:DFT_Ag110}(a)} shows the model for the calculation of a single H$\_2$THPP molecule adsorbed on Ag(110) with the indicated azimuthal rotation $\varepsilon$ of the molecule relative to $[1\bar{1}0]$. After the relaxation, H$\_2$THPP shows on Ag(110) a saddle-shape deformation with average angles of the phenyl planes of $\phi\_{ph, ad} = \SI{31}{\degree}$ and of the pyrrole planes of $\phi\_{pyr, ad}^{up} = \SI{23}{\degree}$ and $\phi\_{pyr, ad}^{down} = \SI{22}{\degree}$ relative to the surface plane, similar to the adsorption on Au(111).

The higher corrugation of the (110) surface compared to (111) results in specific energetic favorable and unfavorable adsorption sites. To evaluate the most favorable adsorption geometries, potential energy curves as a function of $\varepsilon$ for different adsorption sites at the distance of $d = \SI{3.2}{\angstrom}$ (average optimal adsorption height) were calculated [Fig.~\hyperref[fig:DFT_Ag110]{\ref*{fig:DFT_Ag110}(d),(e)}].
Compared to Au(111), the adsorption of H$\_2$THPP on Ag(110) is significantly stronger with higher absolute values of the adsorption energies and a smaller molecule--surface separation. Thus, this corresponds to strong physisorption, which clearly shows the known higher reactivity of the Ag(110) surface. Additionally, energy barriers for molecular diffusion and rotation, which can be estimated from Fig.~\hyperref[fig:DFT_Ag110]{\ref*{fig:DFT_Ag110}(d),(e)}, are significantly higher than on Au(111).

For H$\_2$THPP, the intermolecular interactions determine the close-packed two dimensional structure. The molecular arrangement is then rotated relative to the substrate lattice, so that the molecules with orientations $\varepsilon_1$ and $\varepsilon_1 + \SI{90}{\degree}$ have in total the highest absolute adsorption energy.
The two angles of rotation of the molecules relative to $[1\bar{1}0]$ found by STM were $\varepsilon_1 = \SI{30\pm 7}{\degree}$ and $\varepsilon_2 = \SI{-60\pm 7}{\degree}$ which is identical to $+\SI{60}{\degree}$ due to the symmetry of the system. At both angles [marked in Fig.~\hyperref[fig:DFT_Ag110]{\ref*{fig:DFT_Ag110}(d),(e)}], adsorption with the center in between adjacent Ag rows (position 1 and 4) led to good adsorption energies, whereas the positions on top of an Ag row (2 and 5) were the most unfavorable. Often intermediate energies were calculated for the molecular center above the slope of an Ag row (3 and 6).
With this, the formation of the observed structures with gaps between the molecular blocks or stripes can now be explained. For the smallest unit (4 or 6 molecules) the interaction with the substrate is optimized through its rotation relative to $[1\bar{1}0]$. However, if the periodicity of the close-packed arrangement would be continued [as it is on Au(111)], some molecules would need to adsorb on the unfavorable positions on top of Ag rows, which are avoided.
The model for the epitaxy of the stripe-like structure in Fig.~\hyperref[fig:streifen_Ag110]{\ref*{fig:streifen_Ag110}(b)} shows that consequently three molecules is the maximum favorable width, whereby in the unit cell one position in the middle between adjacent Ag rows and two on the slope of Ag rows are occupied by molecules.
In the direction along the molecular stripe, molecules in each of the three rows are located on similar sites but likely incommensurately, e.g. all molecules of the central row ($\varepsilon_1 \approx \SI{30}{\degree}$) with the center in between Ag rows.
In the structures consisting of small molecular blocks, the discontinuation in two directions allows for the occupation of in average more energetically favorable adsorption sites. As can be seen in the model of this coincident epitaxy in Fig.~\hyperref[fig:struktur_Ag110]{\ref*{fig:struktur_Ag110}(b)}, in the unit cell with six molecules half can adsorb with the center in between Ag rows and the other half on the less favorable slope of Ag rows. However, this increased interaction with the substrate comes at the cost of less hydrogen bonds per molecule compared to the stripe-like structure, or if viewed in the other way: The higher intermolecular interaction in the stripe-like structure compensates for the adsorption on more positions with lower $|E\_{ads}|$.

The most favorable adsorption energies were found for $\varepsilon = \SI{0}{\degree}$ or small angles of rotation relative to $[1\bar{1}0]$ and the single molecule adsorbed with the center above a silver row of the Ag(110) substrate (position 2 and 5). This matches well with H$\_2$THPP in the molecular chains along the $[1\bar{1}0]$ direction found with STM. For the epitaxy of the chains [Fig.~\hyperref[fig:chain]{\ref*{fig:chain}(c)}], this means that the molecules are always located on the best adsorption positions. Thus, in the one-dimensional arrangement intermolecular interactions are reduced the most in favor of the best interaction with the substrate.

The formation of molecular chains was also observed for a Pt-porphyrin with more bulky di-tertiary-butylphenyl-groups on Ag(110).\cite{Yokoyama2008} As explanation for this linear growth behavior, a template-guided adsorption on rails of Ag adatoms was suggested. This could alternatively explain why our molecular chains of H$\_2$THPP were mostly located near step edges because it is known that on Ag(110) at \SI{300}{\kelvin} Ag atoms easily detach from step edges and are able to diffuse over the surface.\cite{Morgenstern1999, Zambelli1998} H$\_2$THPP molecules could then arrange over these atomic rails or capture and stabilize the Ag atoms underneath.
Using DFT, the energetic minimum for the relaxed adsorption on an additional row of Ag atoms [Fig.~\hyperref[fig:DFT_Ag110]{\ref*{fig:DFT_Ag110}(c)}] was found at $E\_{ads} = \SI{-6.24}{eV}$, which is indeed more favorable than the adsorption on the normal Ag(110) at the same molecular orientation ($\varepsilon = \SI{0}{\degree}$). The small distance of \SI{2.3}{\angstrom} between the nitrogen and silver adatoms indicates a strong interaction close to chemisorption. This would corroborate the suggestion of underlying Ag adatoms. However, a single rail of adatoms cannot explain the always observed, small, alternating displacement of the molecules vertical to the chain. Also, clear experimental evidence would be needed to prove this special adsorption geometry for the here studied system.

\section{Conclusions}

5,10,15,20-tetra(\textit{p}-hydroxyphenyl)porphyrin molecules form highly ordered hydrogen-bonded structures in the molecular adsorbate layer.
At submonolayer coverage on Au(111), the molecules self-assemble into a square arrangement characterized by pair-wise O$-$H$\cdots$O hydrogen bonding. Near a complete monolayer, a second structure with an oblique unit cell, higher packing density and more hydrogen bonds per molecule was found. The coincident epitaxy on Au(111) does not have a major influence on this supramolecular ordering, which is defined by a zigzag-chain-wise hydrogen bonding scheme. This is not the case for the adsorption on the highly anisotropic Ag(110) substrate, where the observed structures of H$\_2$THPP are related to different balances between intermolecular interactions (number of hydrogen bonds) and the occupation of favorable adsorption positions. The strong intermolecular interactions lead to structures with an oblique molecular arrangement and well defined orientation of the saddle-shaped macrocycle similar to the adsorption on Au(111). However, the short-range periodicity of the close-packed arrangement is discontinued and large unit cells are formed because unfavorable adsorption sites on the Ag(110) lattice are avoided.
Furthermore, the growth of molecular chains of H$\_2$THPP along the $[1\bar{1}0]$ direction of Ag(110) was found, whereby the molecule--substrate interaction is the dominating factor.

In summary, we demonstrated the effect of strengthening intermolecular interactions between porphyrin molecules \textit{via} hydrogen bonds, which leads to the formation of novel supra-molecular structures and complex epitaxial behaviors.

\section*{Acknowledgements}

This work has been financially supported by the Deutsche Forschungsgemeinschaft (DFG) through the Research Unit FOR 1154 and the Fonds der Chemischen Industrie (FCI).
Computational resources were provided by the ``Chemnitzer Hochleistungs-Linux-Cluster'' (CHiC) at the Technische Universit\"at Chemnitz.

\section*{References}

\bibliographystyle{model1a-num-names}
\bibliography{THPP_struc}

\end{document}